\documentclass[
    prl,twocolumn,
    superscriptaddress,
    amsmath,amssymb,aps,
    floatfix
]{revtex4-1}
\usepackage{graphicx}
\usepackage{times}
\usepackage[varg]{txfonts}
\usepackage[
    colorlinks=true, urlcolor=blue,
    linkcolor=blue, citecolor=blue
]{hyperref}% add hypertext capabilities
\usepackage[mathlines]{lineno}
\usepackage{braket}
\usepackage{floatrow}
\usepackage[caption=false]{subfig}
\usepackage[version=4]{mhchem}
\usepackage[english]{babel}
\usepackage[autostyle, english = american]{csquotes}
\MakeOuterQuote{"}
\graphicspath{
    {./images/}
}

\begin{document}
%\preprint{APS/123-QED}
\title{Pseudogap phase as fluctuating pair density wave}

\author{Zheng-Yuan Yue}
\thanks{These authors contribute equally.}
\affiliation{Department of Physics, The Chinese University of Hong Kong, Sha Tin, New Territories, Hong Kong, China
}%

\author{Zheng-Tao Xu}
\thanks{These authors contribute equally.}
\affiliation{State Key Laboratory of Low Dimensional Quantum Physics and Department of Physics, Tsinghua University, Beijing 100084, China}

\author{Shuo Yang}
\email{shuoyang@tsinghua.edu.cn}
\affiliation{State Key Laboratory of Low Dimensional Quantum Physics and Department of Physics, Tsinghua University, Beijing 100084, China}
\affiliation{Frontier Science Center for Quantum Information, Beijing 100084, China}
\affiliation{Hefei National Laboratory, Hefei 230088, China}

\author{Zheng-Cheng Gu}
\email{zcgu@phy.cuhk.edu.hk}
\affiliation{Department of Physics, The Chinese University of Hong Kong, Sha Tin, New Territories, Hong Kong, China
}%

% but any date may be explicitly specified

\begin{abstract}
    The physical nature of pseudogap phase is one of the most important and intriguing problems towards understanding the key mechanism of high temperature superconductivity in cuprates.
    Theoretically, the square-lattice $t$-$J$ model is widely believed to be the simplest toy model that captures the essential physics of cuprate superconductors. 
    We employ the Grassmann tensor product state approach to investigate uniform states in the underdoped ($\delta \lesssim 0.1$) region. 
    In addition to the previously known uniform $d$-wave state, we discover a strongly fluctuating pair density wave (PDW) state with wave vector $Q = (\pi, \pi)$. 
    This fluctuating PDW state weakly breaks the $C_4$ rotational symmetry of the square lattice and has a lower or comparable energy to the $d$-wave state (depending on doping and the $t/J$ ratio), making it a promising candidate state for describing the pseudogap phase. 
\end{abstract}

\maketitle

\paragraph*{Introduction ---} 
High-temperature superconductivity in cuprate materials remains a challenging field in condensed matter physics, even decades after its discovery \cite{Bednorz1986}. 
Cuprates are Mott insulators at half-filling, but they become superconducting upon doping \cite{Lee2006RMP}. 
The underdoped region of the cuprate phase diagram is particularly intriguing due to the emergence of the pseudogap below a certain temperature $T^*$ \cite{Timusk1999, Keimer2015}. It is now shown to be associated with the breaking of various symmetries, including the $C_4$ rotational symmetry of the $\ce{CuO_2}$ planes \cite{Daou2010, Lawler2010, Wu2017, Sato2017, Murayama2019}. 
Interestingly, the model compound $\ce{HgBa_2CuO_{4+\delta}}$ (Hg1201, which has a tetragonal structure and only one $\ce{CuO_2}$ plane per unit cell) \cite{Barisic2008} exhibits nematicity below $T^*$ along the diagonal directions $[110]$ or $[1\bar{1}0]$ \cite{Murayama2019}. 
Such a rotational symmetry breaking is very different from the charge-density-wave (CDW) phase, which breaks translational symmetry. 
In the pseudogap phase, there is growing interest in the possible existence of a pair density wave (PDW) order \cite{fradkin2015colloquium, agterberg2020pdw}. PDW is characterized by the formation of Cooper pairs that carry a nonzero total momentum $Q$, resulting in a spatial modulation of the superconducting order parameter. Numerous experiments support the presence of PDW in cuprate superconductors \cite{Chen2004, Edkins2019, Hamidian2016, Ruan2018, choubey2020pdw, du2020pdw, Li2021, Lee2023}. However, theoretical understanding of the PDW state is based primarily on phenomenological models, and ongoing debates persist regarding its microscopic foundations and its relation to the pseudogap \cite{lee2014amperean, Wang2015, Agterberg2015, Cai2017, Norman2018, dai2018pdw, dai2020pdw, flucPDW2023, Wu2023pdw, Setty2023}.

In this work, we use the Grassmann tensor product state (TPS) method \cite{Gu2010Grassmann, Gu2013Grassmann} (which can also be equivalently formulated in fermionic tensors \cite{Bultinck2017, Mortier2024}) to study the $t$-$J$ model on a square lattice \cite{ZhangRice1988tJ}, which is believed to capture the essential physics of the $\ce{CuO_2}$ plane. 
Previous numerical studies using techniques such as Density Matrix Renormalization Group (DMRG) \cite{White1998, White2000, Jiang2021, Gong2021, Jiang2022, Chen2023, Jiang2023, Lu2024}, infinite Projected Entangled-Pair States (iPEPS) \cite{Corboz2011, Corboz2014, Li2021PEPS}, and Variational Monte Carlo (VMC), which utilizes a $d$-wave slave-boson projective wave function \cite{Himeda2002, Sorella2002, Ivanov2004, Lugas2006, Capello2008, Hu2012} have revealed various competing orders, including a $d$-wave superconducting (SC) order that coexists with antiferromagnetic (AFM) order, as well as partially filled stripe orders with varying periods. 

Here we report the discovery of a new \emph{uniform and fluctuating} PDW state with $Q = (\pi, \pi)$ (coexisting with antiferromagnetic order in the doping range $\delta \lesssim 0.1$). 
Important features of this state include: (a) the nearest-neighbor (NN) singlet pairing vanishes as the bond dimension $D \to \infty$, implying strong quantum fluctuation; (b) the hole density, staggered magnetization, and the NN singlet pairing magnitude are uniform; and (c) the $C_4$ rotational symmetry is weakly broken to mere reflections about the diagonal next-nearest-neighbor (NNN) bonds. These contrast sharply with previously found PDW states, which are essentially spatially modulated $d$-wave states (also called "antiphase" $d$-wave states) with a smaller but nonzero wave vector and a static singlet pairing pattern \cite{Himeda2002, Berg2009, Yang2009, agterberg2020pdw, Chen2023}. 
Our fluctuating PDW state has a lower or comparable energy to the uniform $d$-wave state at low doping (depending on doping and the $t/J$ ratio).
%This state also manifests weaker SC and AFM orders compared to the $d$-wave state, indicative of strong quantum fluctuations. 
These features suggest that the pseudogap phase may actually be a distinct quantum state, which strongly competes with the $d$-wave superconducting state at low doping.
%characterized by the fluctuating pair density wave.

\paragraph*{Model and methods ---}\label{sec:methods}
The Hamiltonian of the well-known $t$-$J$ model reads
\begin{equation}
    H = -t\sum_{\langle ij\rangle , \sigma}\left(
        \tilde{c}_{i\sigma}^\dagger
        \tilde{c}_{j\sigma} + h.c.
        \right) + J\sum_{\langle ij\rangle} \left(
        \mathbf{S}_i \cdot \mathbf{S}_j
        - \frac{1}{4} \hat{n}_i \hat{n}_j
    \right),   
\end{equation}
Here $\braket{ij}$ sums over NN bonds.
$
    \tilde{c}_{i\sigma} 
    = c_{i\sigma}(
        1 - c^\dagger_{i\bar{\sigma}}
        c_{i\bar{\sigma}}
    )
$
is the electron operator defined in the no-double occupancy subspace. 
$
    \hat{n}_i = \sum_\sigma c_{i\sigma}^{\dagger}c_{i\sigma}
$ 
is the electron number operator. 
$
    \mathbf{S}_i = (1/2) \sum_{\alpha,\beta}
    c^\dagger_{i\alpha}
    \boldsymbol{\sigma}_{\alpha\beta} c_{i\beta}
$ 
is the spin-1/2 operator, with $\boldsymbol{\sigma}$ being Pauli matrices. 
The ground state $\ket{\psi}$ is taken as a two-dimensional translational invariant Grassmann TPS \cite{Gu2013Grassmann}, which is divided into two sub-lattices $A$ and $B$, each generated by local Grassmann tensors $T_A$ or $T_B$. 
There are also Schmidt weights $\Lambda$ on the four types of bonds (named $1$ to $4$), which are diagonal dual Grassmann matrices.
% used to approximate the environment surrounding a local tensor in the simple update (SU) scheme.
\begin{equation}
    \ket{\psi} = \begin{matrix}
        \includegraphics[scale=0.55]{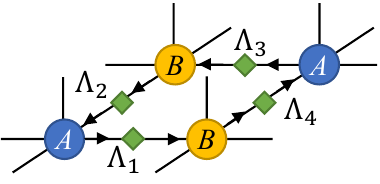}
    \end{matrix}.
\end{equation}
The arrow on each bond indicates the order of the two dual Grassmann numbers in the Grassmann metric used by the contraction \cite{Gu2013Grassmann}. 
All virtual indices have the same total dimension $D$ and the bosonic (even-parity) subspace dimension $D_e$. 
The physical index is $3$ dimensional, labeled as $\ket{\uparrow}$ (spin-up), $\ket{\downarrow}$ (spin-down) and $\ket{0}$ (empty site). 
The double occupancy state is excluded. 
We adopt the slave fermion point of view \cite{frohlich1992} and assign $\ket{\uparrow}, \ket{\downarrow}$ to the bosonic subspace. 

Starting from an initialization of $\ket{\psi}$, the ground state is obtained using the simple update (SU) method, which is an imaginary time evolution algorithm \cite{vidal2007simpleupdate,jiang2008accurate,supp}.
Since nonuniform states (e.g. with stripe order or charge density wave order) break the translational symmetry, they are irrelevant to pseudogap physics. Furthermore, a previous study with the full update (FU) scheme \cite{Corboz2014} suggests that the energy of uniform states may eventually be comparable to that of nonuniform states as $D$ increases, especially with smaller doping. DMRG also prefers uniform SC state on wider systems \cite{Jiang2023, Lu2024}. Therefore, we focus on uniform states, which are obtained by averaging the Schmidt weights after updating them all once. 
Doping $\delta$ is controlled by adding a chemical potential term $-\mu \sum_i \hat{n}_i$ ($\mu > 0$) to the Hamiltonian, and a larger $\mu$ leads to smaller doping. 
The evolution time step slowly decreases to ensure its convergence. 
The update stops when the change in the Schmidt weight is sufficiently small. 
We measure the resulting ground state with the variational uniform matrix product state (VUMPS) method \cite{fastcontract2018, tangentspace2019, vumps2018}.
The measurement result converges when the dimension of the MPS virtual boundary bond is $\chi \gtrsim 4D$ \cite{supp}. 

\begin{figure}[tb]
    % \centering
    \includegraphics[scale=0.85]{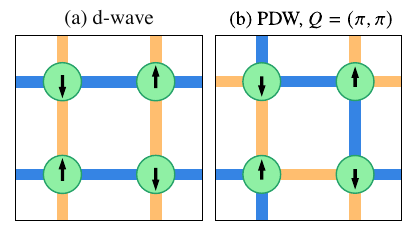}
    \caption{
    NN singlet pairing $\Delta_{ij}$ patterns of (a) the $d$-wave state, and (b) the fluctuating PDW state with $Q = (\pi,\pi)$. 
    Both states are uniform and 
    $|\Delta_{ij}|$ on each NN bond has the same magnitude. 
    Its sign is shown by the bond colors (light orange for $\Delta_{ij} > 0$ and dark blue for $\Delta_{ij} < 0$). 
    In both states, we observe co-existing anti-ferromagnetic order at small doping, i.e. $\braket{\mathbf{S}_{i \in A}} = -\braket{\mathbf{S}_{i \in B}}$.}
    \label{fig:config}
\end{figure}

\begin{figure}[tb]
    % \centering
    \includegraphics[scale=0.65]{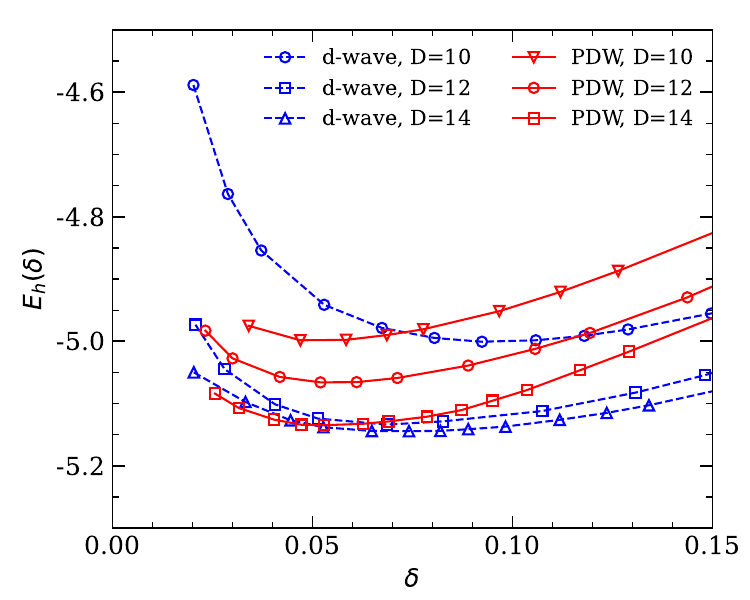}
    \caption{
    Energy per hole $E_h$ for the $d$-wave and the PDW states at $t/J = 3.0$, measured with $\chi = 64$. 
    }
    \label{fig:ehole_vs_doping}
\end{figure}

\begin{figure}[tb]
    % \centering
    \includegraphics[scale=0.55]{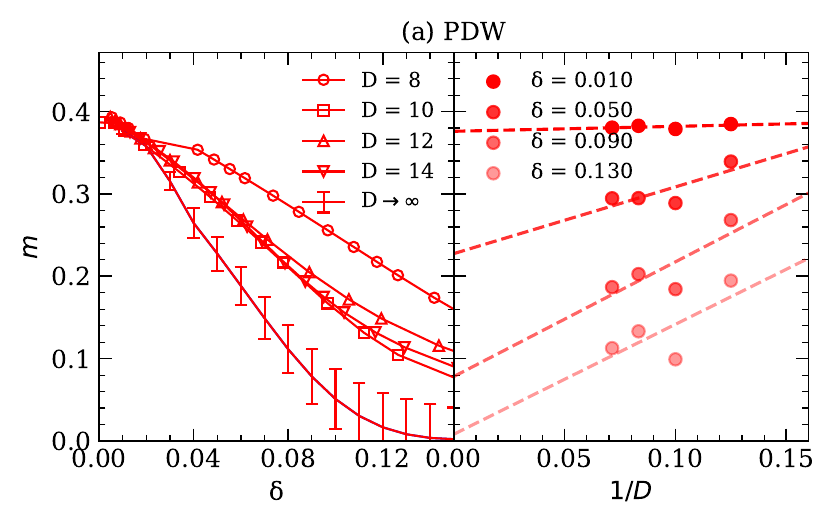}
    \\
    \includegraphics[scale=0.55]{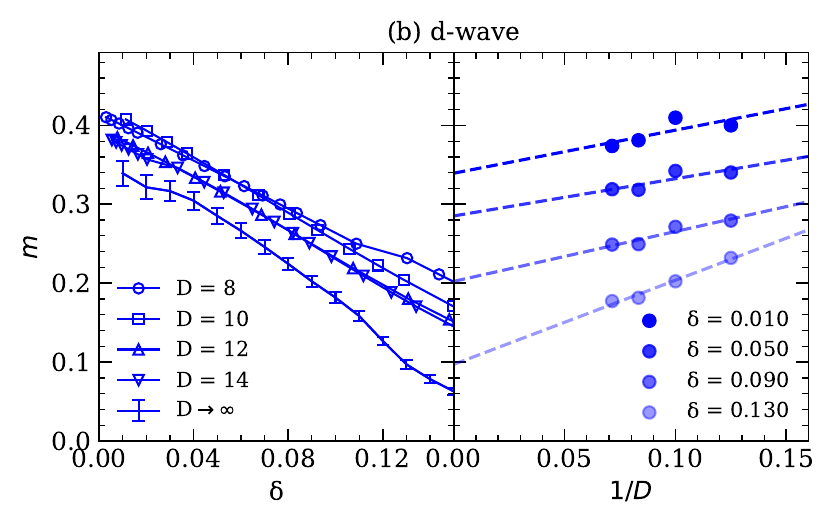}
    \caption{
    $1/D$ scaling of staggered magnetization $m$ at $t/J = 3.0$ for (a) fluctuating PDW and (b) $d$-wave states, measured with $\chi = 64$. Vertical bars on the $D \to \infty$ curves are linear fit errors. }
    \label{fig:vsD-mag-t3}
\end{figure}

\begin{figure}[tb]
    \includegraphics[scale=0.55]{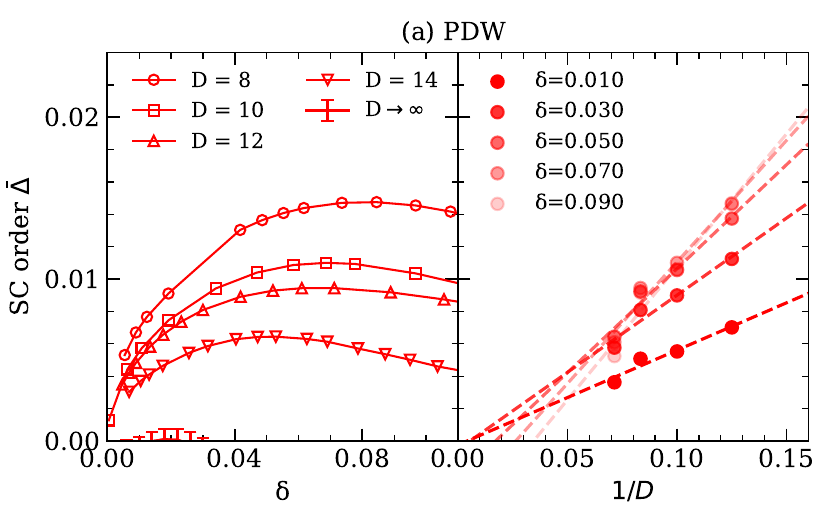}
    \\
    \includegraphics[scale=0.55]{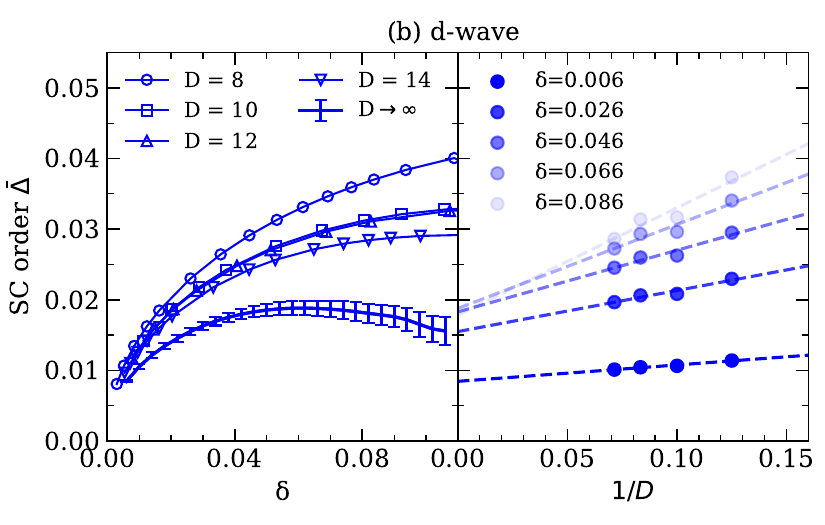}
    \caption{
    $1/D$ scaling of average magnitude of NN singlet pairing $\bar{\Delta}$ at $t/J = 3.0$ for (a) fluctuating PDW and (b) $d$-wave states, measured with $\chi = 64$. 
    Within the fitting error, $\bar{\Delta}$ of the fluctuating PDW state vanishes as $D \to \infty$. }
    \label{fig:vsD-sc-t3}
\end{figure}

\paragraph*{Uniform competing ground states ---}\label{sec:results} 
Starting from random initialization, we find that the ground state converges to two types of uniform state: the well-known $d$-wave SC state (Fig. \ref{fig:config}(a)), and a new PDW state (Fig. \ref{fig:config}(b)). Here, the SC order is detected by the real space singlet pairing amplitude on each bond:
$
    \Delta_{ij} = \braket{
        c_{i\uparrow} c_{j\downarrow}
        - c_{i\downarrow} c_{j\uparrow}
    } / \sqrt{2}.
$
In the $d$-wave state, the singlet pairing pattern is
$
-\Delta_{(x,y)(x+1,y)}
= \Delta_{(x,y)(x,y+1)} \equiv \Delta
$. 
In the PDW state, 
$
\Delta_{(x,y)(x+1,y)}
= -\Delta_{(x,y)(x,y+1)}
\equiv (-1)^{x+y} \Delta
$, 
which varies periodically along the diagonal direction of the lattice, with wave vector $Q = (\pi, \pi)$. We note that the hole density at each site is uniform, and the magnitude $|\Delta_{ij}|$ is the same on all NN bonds. 
This is rather different from PDW orders in the existing literature, which are spatially modulated $d$-wave states ($|\Delta_{ij}|$ varies periodically), coexisting with charge or spin density waves. Strictly speaking, in the presence of AFM order, the momentum $Q = (\pi, \pi)$ is identified with $Q = (0, 0)$ due to unit cell doubling. However, the PDW state discovered here can still survive in the absence of a long-range AFM order, e.g., with a bigger $t/J$ ratio or at larger doping. Therefore, we believe that the coexistence of an AFM order does not change the physical nature of such a PDW state.

We also measured the singlet pairing $\Delta_{ij}$ in further neighbors. Interestingly, for both $d$-wave and PDW states, $\Delta_{ij} = 0$ (up to numerical errors) on NN bonds. Furthermore, all same-sub-lattice SC pairings (along $x$ and $y$ directions) vanish in the PDW state, that is, $\Delta_{i, i+n\hat{x}} = \Delta_{i, i+n\hat{y}} = 0$ when $n$ is an even number (see Supplemental Materials \cite{supp} for details). Such a super-selection rule for SC pairings indicates a hidden projective symmetry group (PSG) that has never been realized by a projective mean-field wave function before.

To better visualize the energy difference, we plot the energy per hole $E_h = (E_s - E_0)/\delta$ (in units of $J$) in Fig. \ref{fig:ehole_vs_doping}. 
Here, $E_s$ is the energy per site, and $E_0 = -1.169438$ is its value at zero doping \cite{sandvik1997heis}. 
The energies for the two states at $t/J = 3$ with various $D$ are quite close throughout the region $0 < \delta \lesssim 0.1$. 
All of these energies are much lower than the previous VMC result \cite{Ivanov2004}, but slightly higher than the stripe states in the previous study for doping $\delta > 0.05$ (see Supplemental Materials \cite{supp}). 

The magnetic order detected by magnetization on each site is defined as
$
m_i = \sqrt{
    \braket{S^x_i}^2 
    + \braket{S^y_i}^2 
    + \braket{S^z_i}^2
}
$. 
With finite $D$, both states have nonzero staggered magnetization $m$. To confirm the existence of an AFM order, we perform $1/D$ scaling over $m$ (linear fit with respect to $1/D$), and we find that $m$ is still nonzero as $D \to \infty$ (see Fig. \ref{fig:vsD-mag-t3}) at small doping. 
We note that SU overestimates the magnetic order: while $m$ in the $d$-wave state is always nonzero in the whole underdoped region $\delta \lesssim 0.15$, FU suggests that $m$ vanishes around $\delta \sim 0.1$ \cite{Corboz2014}. 
Meanwhile, our SU result shows that $m$ drops to 0 for the PDW state at $\delta \simeq 0.12$.
However, we believe that FU or other global optimization methods will further reduce the magnetization in the PDW state as well. 

\paragraph*{Fluctuating pair density wave and weak $C_4$ rotational symmetry breaking ---} 
%By increasing bond dimension D, we can introduce more quantum fluctuations into the Grassmann TPS ansatz.
Fig. \ref{fig:vsD-sc-t3} shows the magnitude of singlet pairing $|\bar{\Delta}|$ on NN bonds, obtained with various bond dimensions $D$. 
%In both states, $m$ and $|\bar{\Delta}|$ are nonzero at small doping, indicating co-existing AFM and SC orders. 
$|\bar{\Delta}|$ is generally smaller in the PDW state than in the $d$-wave state at the same doping.
Furthermore, $|\bar{\Delta}|$ of the PDW state eventually vanishes in the $D \to \infty$ limit (within the fit error) after the $1/D$ scaling. 
This is in contrast to the $d$-wave state, where a finite pairing amplitude always survives after $1/D$ scaling. 
We conjecture that such a fluctuating PDW state could naturally explain the experimentally observed pseudogap phase in cuprate materials \cite{dai2020pdw, flucPDW2023}. 
In particular, the fluctuating PDW state suggests that the pseudogap phase should be regarded as a zero-temperature quantum phase with local pairing but without global phase coherence. 
%Our results also resolve the long-standing puzzle why $d$-wave superconductivity is always suppressed by at very low doping in cuprate materials while all previous VMC and iPEPS studies always predict the coexistence of the AFM order and the $d$-wave SC order for the $t$-$J$ model at low doping. 

\begin{figure}[tb]
    \includegraphics[scale=0.65]
    {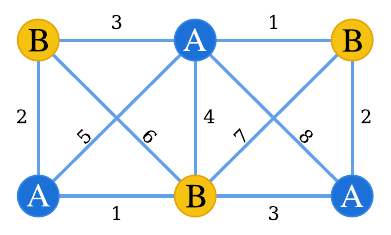}
    \caption{
    Non-equivalent NN bonds (1 to 4) and NNN bonds (5 to 8) under bipartite lattice translation.}
    \label{fig:nb-bonds}
\end{figure}

\begin{figure}[tb]
    \includegraphics[scale=0.6]{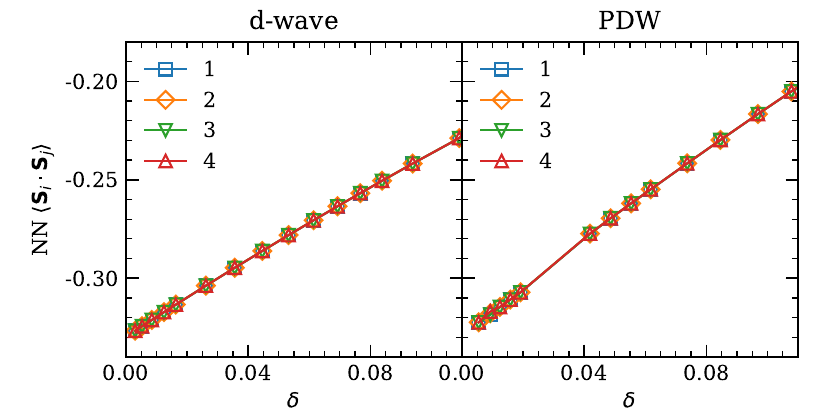}
    \\
    \includegraphics[scale=0.6]{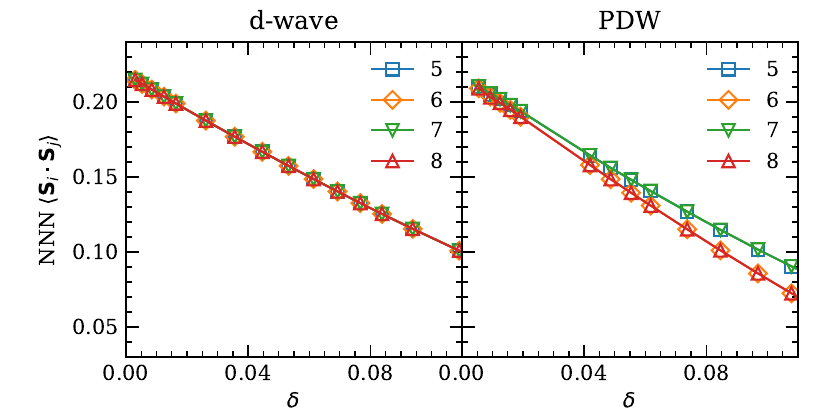}
    \caption{
    Spin correlation $\braket{\mathbf{S}_i \cdot \mathbf{S}_j}$ on NN and NNN bonds in $D = 8$ states at $t/J = 3.0$, measured with $\chi = 32$.}
    \label{fig:nb-spincor}
\end{figure}

\begin{figure}[tb]
    \includegraphics[scale=0.6]{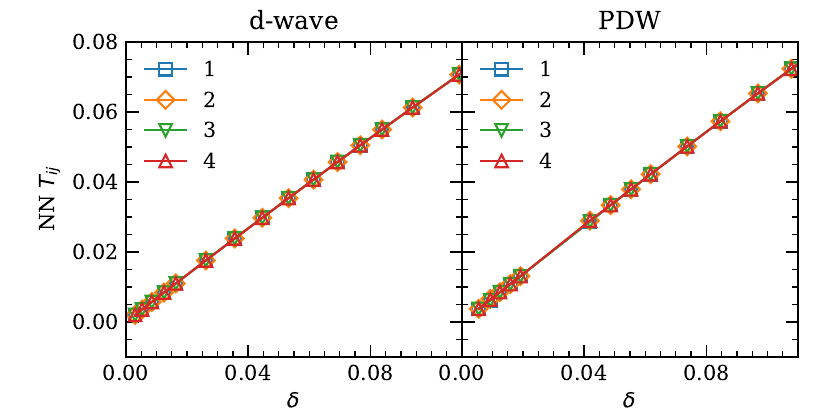}
    \\
    \includegraphics[scale=0.6]{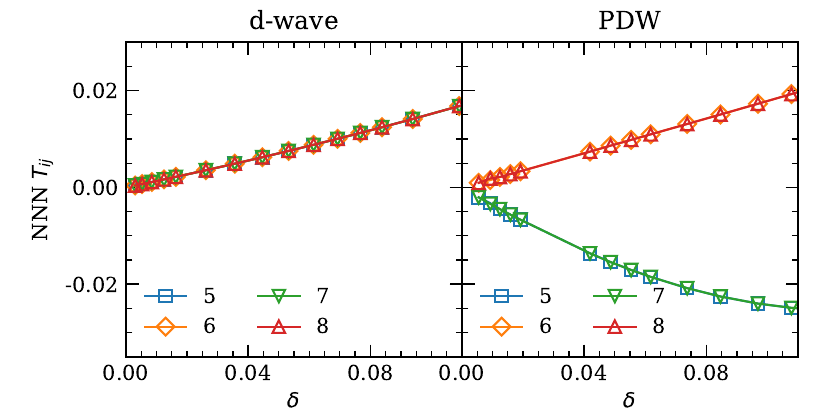}
    \caption{
    Hopping $T_{ij} = \sum_\sigma \braket{c^\dagger_{i\sigma} c_{j\sigma}}$ on NN and NNN bonds in $D = 8$ states at $t/J = 3.0$, measured with $\chi = 32$.}
    \label{fig:nb-hopping}
\end{figure}

The smoking gun evidence to identify the presence of a fluctuating PDW state is weak breaking of the $C_4$ rotational symmetry, which can be detected by measuring the spin-spin correlation $\braket{\mathbf{S}_i \cdot \mathbf{S}_j}$ on nonequivalent NN and NNN bonds (see Fig. \ref{fig:nb-bonds}, labeled from $1$ to $8$). 
The results are shown in Fig. \ref{fig:nb-spincor}.
Despite the bipartite structure of the Grassmann TPS, the spin-spin correlation still has the full lattice translational symmetry in both states. 
However, the PDW state weakly breaks the $C_4$ rotational symmetry, since the correlations on the bonds $5$, $7$ and $6$, $8$ are different.
Reflection symmetry on the two diagonal lines still survives, which might provide a natural explanation for the diagonal nematicity observed in the pseudogap phase of Hg1201 \cite{Murayama2019}.
Symmetry breaking becomes more evident as $\delta$ increases, as shown by the increased discrepancy between the spin-spin correlation in two diagonal directions. 
In contrast, the spin-spin correlation in the $d$-wave state maintains the full lattice symmetry.
%regardless of the value of $\delta$. 

The $C_4$ symmetry breaking is also revealed by the NNN hopping $T_{ij} \equiv \sum_\sigma \braket{c^\dagger_{i\sigma} c_{j\sigma}}$. As seen in Fig. \ref{fig:nb-hopping}, the sign of NNN hopping is the same as the sign of NN hopping in the $d$-wave state, while in the PDW state, $T_{ij}$ on NNN bonds along the two diagonal directions have different signs and magnitudes. However, reflection symmetry on the two diagonal lines still survives. 
Therefore, the $d$-wave state can be further stabilized over the fluctuating PDW state by adding an NNN hopping term 
$
-t' \sum_{\langle\langle i,j \rangle\rangle,\sigma} 
(c^\dagger_{i\sigma} c_{j\sigma} + h.c.)
$
to the Hamiltonian with $t' > 0$. 
This agrees with previous DMRG results that $t' > 0$ enhances the $d$-wave SC order \cite{Jiang2021, Jiang2022, Chen2023, Lu2024}. 
However, in realistic hole-doped cuprate materials with $t'<0$, we believe that such a fluctuating PDW state should become more stable than the $d$-wave state at low doping. 
This explains why the pseudogap phase was observed only in all hole-doped materials but not on the electron-doped side. 
%Essentially, the fluctuating PDW state can be regarded as the "normal" state for hole-doped materials, and the emergence of a $d$-wave SC state from such an exotic "normal" state might hold the key mechanism of high-temperature superconductivity. 

\begin{figure}[bt]
    % \centering
    \includegraphics[scale=0.65]{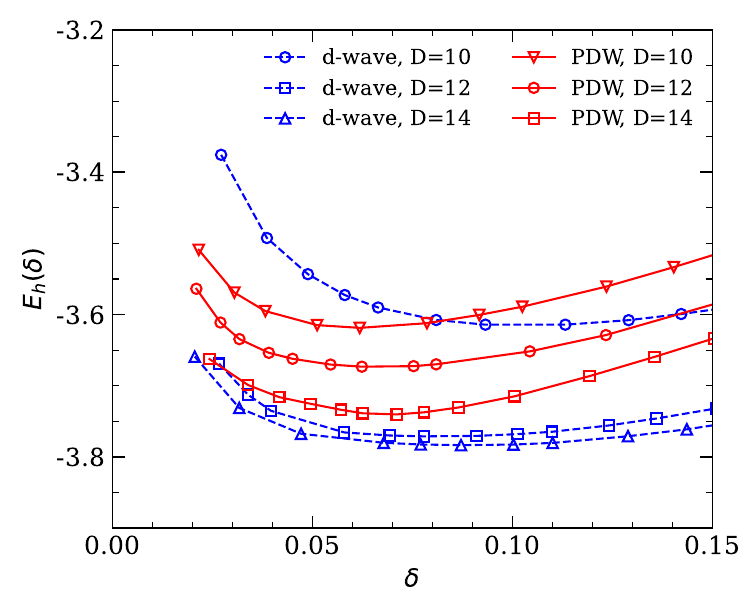}
    \caption{
    Energy per hole $E_h$ of the fluctuating PDW and the $d$-wave states at $t/J = 2.5$, measured with $\chi = 64$.}
    \label{fig:e_vs_doping-2.5}
\end{figure}

\paragraph*{The effect of $t/J$ ratio ---}
Finally, to explore the possible $t/J$ dependence on our result, we repeat the same calculations for $t/J = 2.5$. 
The SC order $\Delta_{ij}$ in the PDW state still exhibits a strongly fluctuating characteristic, which reduces to a very small value as $D \to \infty$, and a similar weak $C_4$ symmetry breaking feature is also observed in NNN bonds (see Supplemental Materials \cite{supp} for details). 
However, the $d$-wave state is now energetically preferred in a wider doping range, but the energy difference between the two states is still small, especially close to half-filling (see Fig. \ref{fig:e_vs_doping-2.5}).
This suggests that the fluctuating PDW state may not only be competitive with $d$-wave state in a limited range of $t/J$, but is actually a generic feature for the doped Mott-insulator on a square lattice. In general, a larger $t/J$ ratio will enhance the quantum fluctuations and stabilize the fluctuating PDW state.
The global phase diagram of $t$-$J$ model can be more accurately determined with further optimization of the SU result using a more sophisticated algorithm, such as FU or gradient-based algorithms. 
Experimentally, our results also indicate that a large $J$ (corresponding to a smaller charge transfer gap in cuprates) might stabilize the $d$-wave superconductivity. 
%and lead to a higher $T_c$.

\paragraph*{Summary and outlook ---} 
\label{sec:summary}
Using the Grassmann TPS approach, we have discovered a uniform PDW state that represents a novel perspective on the emergence of PDW order in cuprates, distinct from the spatially modulating $d$-wave state-related phenomena previously studied. This PDW state exhibits strong quantum fluctuations, as evidenced by the vanishing of the singlet pairing $\Delta_{ij}$ in the limit of large TPS bond dimension. In regions of low doping where uniform states are favored over nonuniform states, our fluctuating PDW state competes strongly with the uniform $d$-wave state. In particular, the weak breaking of the rotational symmetry of the $C_4$ lattice in the fluctuating PDW state makes it a promising candidate for the pseudogap phase. Although all of our results are based on SU, previous tensor network simulations on the $t$-$J$ model suggest that FU will not lead to qualitative differences. 

Finally, the spin-charge separation scenario suggests that there might be an intrinsic connection between the fluctuating PDW state with the $(\pi,\pi)$ wave vector and the peak observed in the dynamic magnetic susceptibility at $(\pi,\pi)$ in under-doped cuprates, as reported in previous neutron scattering studies \cite{RossatMignod1991, Mook1993, Fong1995, Fong1997, Dai1999, Dai2001}. 
This peak is recognized as a signature of the pseudogap phase. 
In particular, the observation of a similar enhancement of the magnetic response to AFM fluctuations below $T^*$ in the model compound Hg1201 \cite{Yu2010, Chan2016, Chan2016PRL}, together with the $C_4$ symmetry breaking feature, further consolidates the uniform fluctuating PDW nature of the pseudogap phase. We believe that other longer-range PDW pattern observed in the pseudogap region, such as the $4a_0$ or $8a_0$ pattern\cite{du2020pdw}, can all be induced from the fundamental $2a_0$ pattern proposed here.
Furthermore, we stress that such a uniform fluctuating PDW state, characterized by the absence of the same sub-lattice superconducting pairing and the weak break of rotational symmetry $C_4$, indicates the existence of a novel PSG structure beyond traditional projected wavefunction techniques. 
Investigating the physical nature of such a new state of matter could provide insight into the microscopic origins of the pseudogap phase and potentially uncover the key mechanisms of high-temperature superconductivity.

\paragraph*{Acknowledgement ---}
We thank Wayne Zheng for the helpful discussions. We also thank Shou-Shu Gong and Xin Lu for providing DMRG data. This work is supported by funding from Hong Kong's Research Grants Council (RFS2324-4S02, CRF C7012-21GF and GRF No. 14302021), the National Natural Science Foundation of China (NSFC) (Grant No. 12174214 and No. 92065205), and the Innovation Program for Quantum Science and Technology (Grant No. 2021ZD0302100).

\bibliographystyle{apsrev4-2}
\bibliography{refs}

\pagebreak
\appendix
\begin{widetext}
\section{Simple update of tensor product states}
\label{app:su}

Starting from an arbitrary state $\ket{\psi_0}$, we can approximate the ground state $\ket{\psi}$ of a Hamiltonian $H$ by the imaginary time evolution $\ket{\psi} \approx e^{-\beta H} \ket{\psi_0}$ with a sufficiently large $\beta$. 
The \textit{simple update} algorithm is a low-cost way to compute evolution when $H$ is the sum of operators with single-site and nearest-neighbor interactions $H_{ij}$. 
With Trotter decomposition, $e^{-\beta H}$ can be expanded as the product of many gates $e^{-\epsilon H_a}$ ($a = 1,...,4$) acting on all type-$a$ bonds with small $\epsilon$. 
Below we describe the update of $\ket{\psi}$ with a Trotter gate $e^{-\beta H_1}$ acting on all type-$1$ bonds
\begin{equation}
    e^{-\epsilon H_1} \ket{\psi}
    = \begin{matrix}
        \includegraphics[scale=0.55]{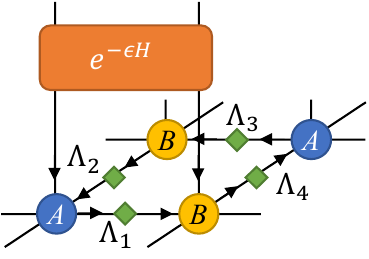}
    \end{matrix}.
\end{equation}

\begin{enumerate}
    \item The weights surrounding the bond to be updated are absorbed into the tensors $A$ and $B$, as an approximation of the environment of the bond. 
    \begin{equation}
        \includegraphics[scale=0.45]{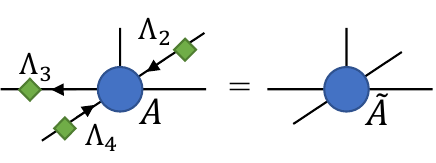},
        \quad
        \includegraphics[scale=0.45]{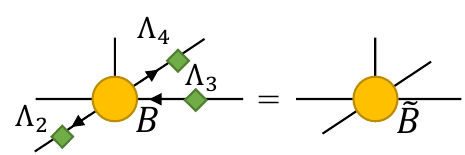}.
        \label{eq:absorb-wt}
    \end{equation}
    
    \item To reduce the computational cost, we first apply QR and LQ decomposition to $\tilde{A}, \tilde{B}$ as
    \begin{equation}
        \includegraphics[scale=0.45]{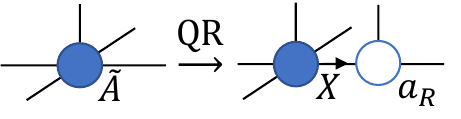},
        \quad
        \includegraphics[scale=0.45]{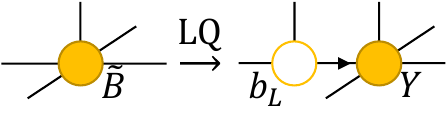}.
    \end{equation}
    The gate $e^{-\epsilon H}$ now acts on the lower-rank tensors $a_R$ and $b_L$ \cite{phien2015fullupdate}
    \begin{equation}
      \begin{matrix}
          \includegraphics[scale=0.5]{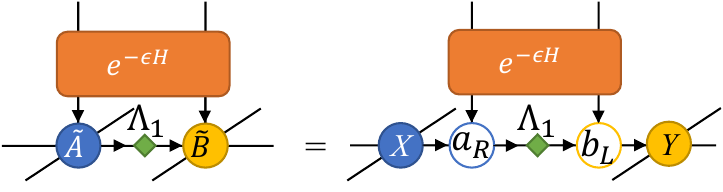}
      \end{matrix}.
    \end{equation}
    Then we perform SVD, obtaining an updated weight $\Lambda_{a}$ (indicated by a prime), $a_R$, and $b_L$ as
    \begin{equation}
      \begin{matrix}
          \includegraphics[scale=0.5]{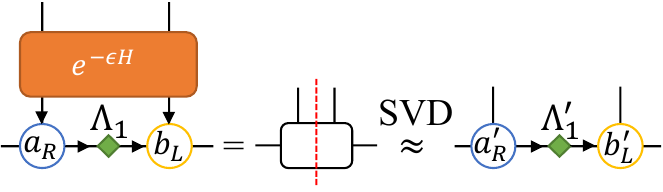}
      \end{matrix}.
    \end{equation}
    
    \item To control the virtual bond dimension, we truncate the weight spectrum $\Lambda$ by keeping only the largest $D$ singular values. 
    Here, $D$ can be different from the virtual bond dimension $D_0$ of the initial state $\ket{\psi_0}$. 
    Note that the singular values of both the even and odd sectors are sorted together. 
    The even (or odd) dimension $D_e$ (or $D_o$) of the virtual index ($D_e+D_o = D$) is the number of these $D$ singular values that come from the even (or odd) sector of $\Lambda$. 
    
    \item The new Schmidt weight is normalized so that the maximum singular value is $1$. 
    The new $a_R, b_L$ tensors are absorbed back into $X, Y$ to produce the new $A, B$ tensors,
    \begin{equation}
      \begin{matrix}
          \includegraphics[scale=0.5]{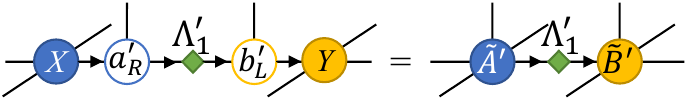}
      \end{matrix}.
    \end{equation}
    The absorbed environment weights are then restored by reversing Eq. \eqref{eq:absorb-wt}. 
    Finally, the updated TPS unit cell is
    \begin{equation}
      e^{-\epsilon H_{a}} \ket{\psi}
      \approx \begin{matrix}
          \includegraphics[scale=0.55]{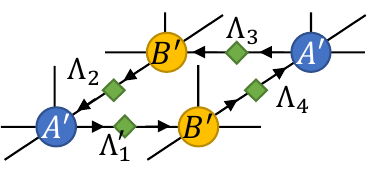}
      \end{matrix}.
    \end{equation}

    \item The steps $1$ to $4$ are repeated for each of the four bonds. 
    To obtain uniform states, we average the four weights after they are all updated once. 
    Otherwise, the system will easily evolve to a non-uniform state. 
    In addition, we only keep states that also have the same $D_e$ on all virtual indices throughout the update process. 
    
    \item The update stops when the change of the averaged weight is sufficiently small, or more precisely,
    \begin{equation}
        \delta \Lambda(n)
        \equiv \frac{1}{D} \sum_{i=1}^D |
            \bar{\Lambda}^{(n)}_i
            - \bar{\Lambda}^{(n-1)}_i
        | \lesssim 10^{-11},
    \end{equation}
    where $\bar{\Lambda}^{(n)}_i$ is the $i$-th averaged weight after the $n$-th round of simple update.
\end{enumerate}

The two main limitations of the simple update method are its difficulty in (a) obtaining states $\ket{\psi}$ with doping $\delta \lesssim 0.02$, since the convergence (decrease in $\delta \Lambda$) becomes very slow and doping $\delta$ becomes very sensitive to the change in $\mu$, and (b) converging from a random initial state when the bond dimension $D$ is large, due to the increase in the number of free parameters in $\ket{\psi}$. 
To address these challenges, we first compute states with $D = 8$ and relatively large doping $0.08 \lesssim \delta \lesssim 0.15$, using a random initialization $\ket{\psi_0}$ with $D_0 = 8$ and even bond dimension $(D_e)_0 = 4$. 
The weights are initialized as identity matrices.
Specifically, all initial states $\ket{\psi_0}$, except those that evolved to half-filling or produce different $D_e$ on each virtual bond during evolution, converge to either $d$-wave or PDW state, both with $D_e = 4$. 
To obtain states with $\delta < 0.08$, we use the previously obtained $d$-wave or PDW state with a close doping level as the initial state to accelerate convergence and avoid evolution toward half-filling. 

To further optimize the energy by increasing $D$, we use states with smaller $D$ as initial states to speed up convergence. 
With the same $D$, it is possible to obtain multiple states with different values of $D_e$.
For example, when initializing with $D = 8$ states to reach $D = 10$, we may obtain the states with $(D_e,D_o) = (4,6)$ or $(5,5)$. 
In such scenarios, we keep the $D_e$ that has the lowest energy. 
For $D$ = 10, 12, 14, we find the states with the lowest energies when $\delta \lesssim 0.1$ have $D_e$ = 5, 5, 6 (PDW) or 7 ($d$-wave) respectively. 
The evolution time step $\epsilon$ is gradually reduced to ensure the convergence of SU.
We start with $\epsilon = 0.01$, and reduce it to $0.001$, $0.0004$, and $0.0001$ as the weight change $\delta\Lambda$ reaches the threshold values $10^{-6}$, $10^{-9}$, and $10^{-10}$, respectively.
For $\delta \lesssim 0.02$, one can begin directly with $\epsilon = 0.001$ or smaller to avoid evolution to half-filling. 

\begin{figure}[bt]
    \includegraphics[scale=0.65]{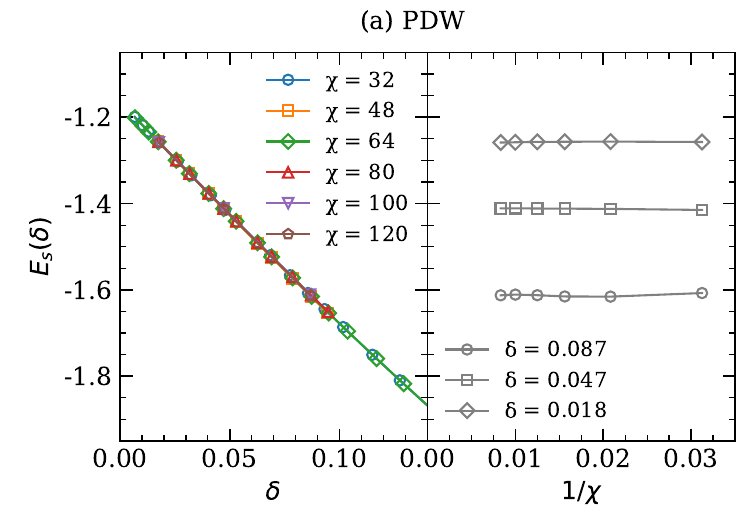}
    \quad
    \includegraphics[scale=0.65]{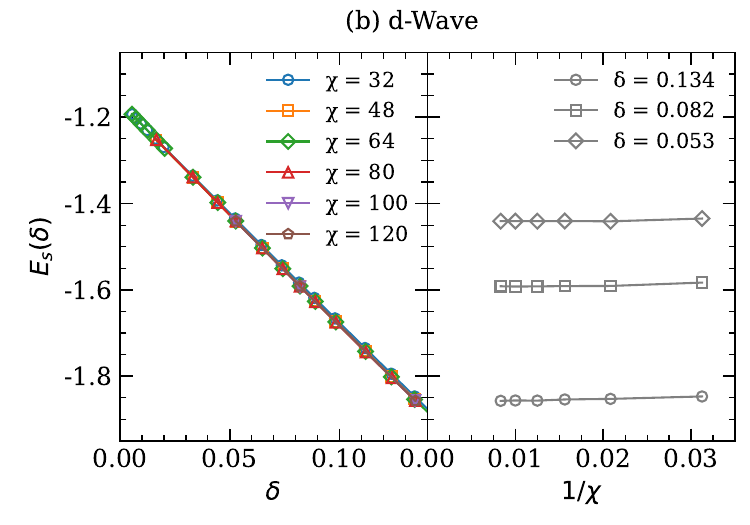}
    \caption{
    Energy per site $E_s$ at $t/J = 3.0$ for (a) fluctuating PDW and (b) $d$-wave states with $D = 14$ measured with different bond dimensions $\chi$ of boundary MPSs (left panels) and their $1/\chi$ scaling (right panels).}
    \label{fig:vschi-esite}
\end{figure}

\begin{figure}[tbh]
    % \centering
    \sidesubfloat[]{\includegraphics[scale=0.6]{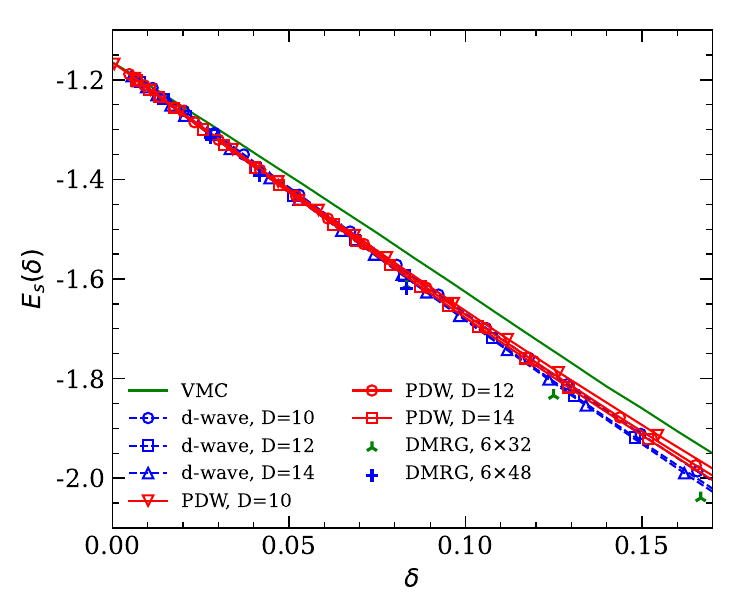}}
    \sidesubfloat[]{\includegraphics[scale=0.6]{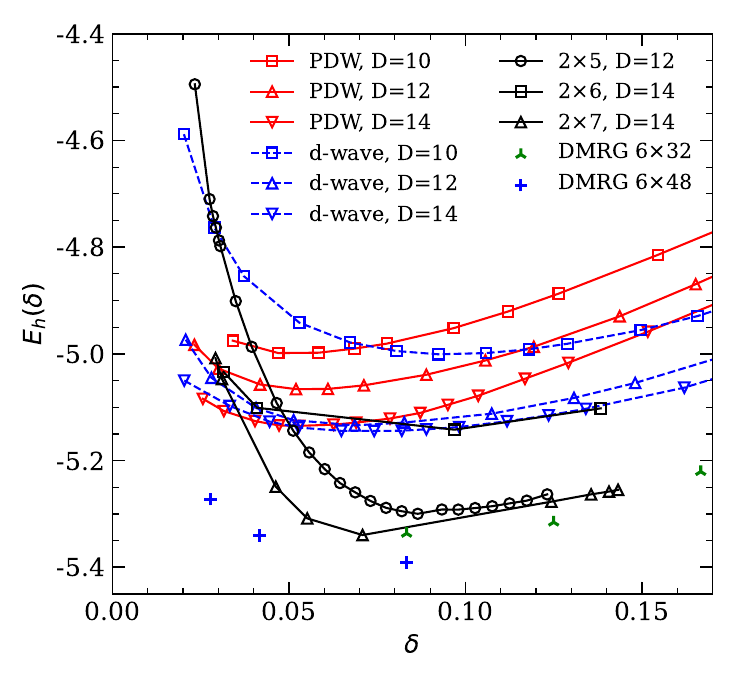}}
    \caption{
        Comparing energies of uniform $d$-wave and fluctuating PDW states with others at $t/J = 3.0$.
        (a) Energy per site $E_s$ comparison with VMC and DMRG (see details in the text). 
        (b) Energy per hole $E_h$ comparison with DMRG and stripe states with unit cell sizes $2 \times 5,\  2 \times 6,\  2 \times 7$ obtained from simple update (see details in the text).
    }
    \label{fig:esite_vs_doping}
\end{figure}

\section{Convergence of VUMPS measurements}
\label{app:details}

To ensure the convergence of VUMPS measurement, we test different boundary MPS virtual bond dimensions $\chi$. 
Fig. \ref{fig:vschi-esite} shows the energy per site $E_s$ for $D = 14$ states measured with various values of $\chi$. 
When $\chi \gtrsim 4D$, the $1/\chi$ scaling is deemed convergent for both the $d$-wave and the fluctuating PDW states, i.e. the change in $E_s$ is of the order $10^{-4}$ if $\chi$ is further increased. 
Thus, we mainly present the $\chi = 64$ data in the main text. 

\section{Comparison with VMC, DMRG and stripe states}

In Fig. \ref{fig:esite_vs_doping}(a), we compare the energy of our $d$-wave and fluctuating PDW states with VMC \cite{Ivanov2004}. For $D \ge 10$, both have energies significantly lower than VMC (especially for larger doping). 
In Fig. \ref{fig:esite_vs_doping}(b), we compare with stripe states with unit cell sizes $2 \times 6,\ 2 \times 7,\ 2 \times 8$ obtained from simple update, which are measured with $\chi = 32$ for $2 \times 5$ and $64$ for the others. 
Although the stripes states have lower energy at larger doping, their $E_h$ increases more rapidly as doping decreases. This trend agrees with previous studies \cite{Corboz2011, Corboz2014}. The uniform states have lower energy at small doping $\delta \lesssim 0.05$. 

We also make a rough comparison with DMRG on 6-leg cylinders \cite{Gong2024dmrg}, which show a stripe state with charge density period $2/(3\delta)$. 
All DMRG energies are calculated with $U(1) \times SU(2)$ symmetry ($U(1)$ is for charge conservation). 
The bond dimensions on $6 \times 32$ and $6 \times 48$ cylinders are $D = 12000$ and 8000, respectively. 
These energies are comparable with the SU results with $2 \times 5$ and $2 \times 7$ unit cells. We note that due to the boundary effect, the DMRG energy per site of the whole cylinder is slightly higher than the energy per site of the bulk stripes.

\section{Singlet pairing on farther neighbors}

We measured $\Delta_{ij}$ on NNN bonds for the $D = 8$ $d$-wave and PDW states with boundary MPS bond dimension $\chi = 32$. 
We find that they are all smaller than $5 \times 10^{-5}$, and can be regarded as zero within the numerical errors. We expect that states with larger $D$ show the same behavior. 
We also measured longer-range $\Delta_{(0,0)(x,0)}$ and $\Delta_{(0,0)(0,y)}$ (see Fig. \ref{fig:singlet-xy}). 
For PDW state, we find that the singlet pairing is zero when $x$ or $y$ is an even number, i.e. the two sites involved are on the same sub-lattice. 

\begin{figure}[tbh]
    \includegraphics[scale=0.65]{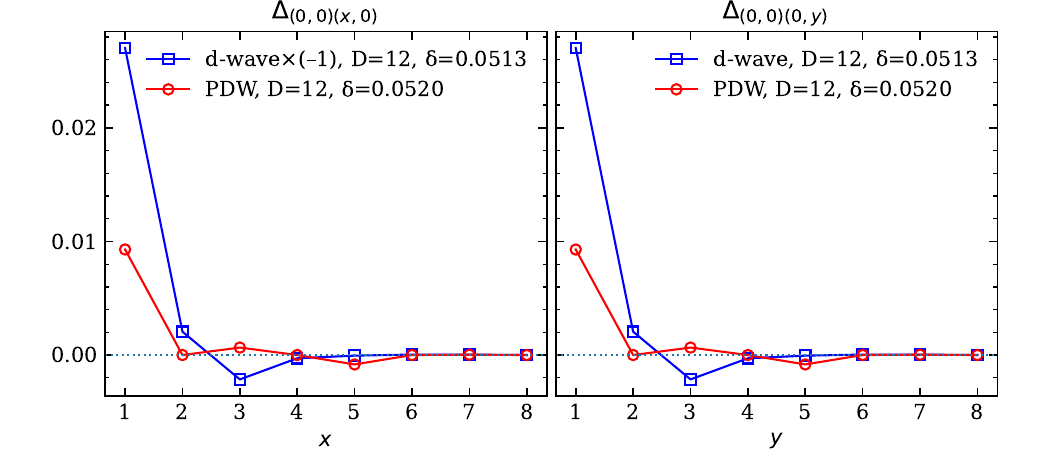}
    \caption{
        Singlet pairing $\Delta_{ij}$ on long-range bonds along $x$ and $y$ directions. $\Delta_{(0,0)(x,0)}$ for $d$-wave states are multiplied by $-1$. The selected $d$-wave state has $D = 5+7$ and doping $\delta = 0.0513$. The selected PDW state has $D = 5+7$ and doping $\delta = 0.0520$. 
    }
    \label{fig:singlet-xy}
\end{figure}

\begin{figure}[tbh]
    % \centering
    \includegraphics[scale=0.6]{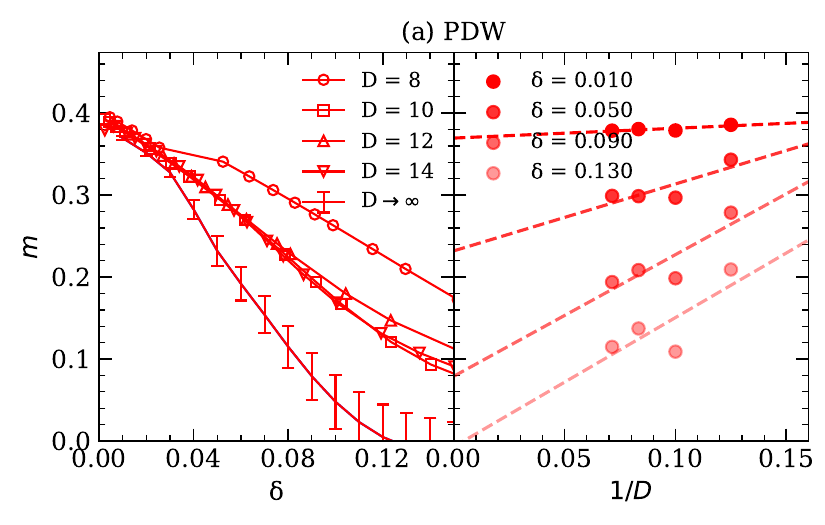}
    \includegraphics[scale=0.6]{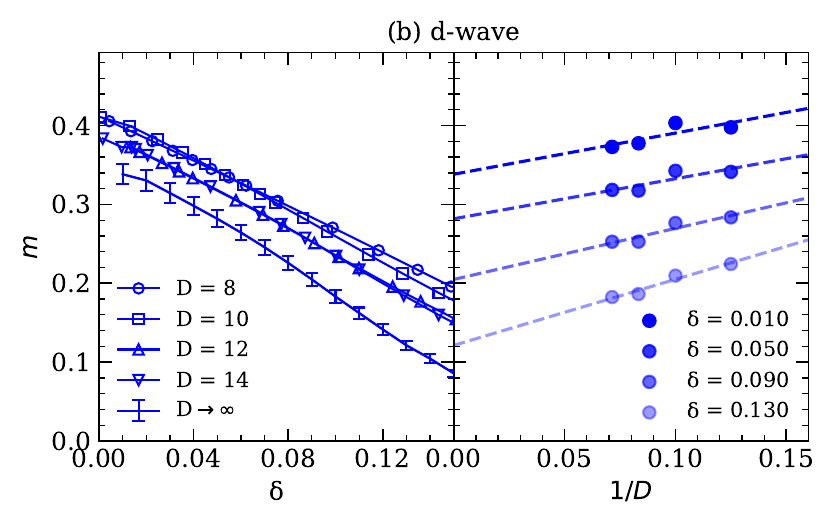}
    \caption{
    $1/D$ scaling of staggered magnetization $m$ at $t/J = 2.5$ for (a) fluctuating PDW and (b) $d$-wave states measured with $\chi = 64$. }
    \label{fig:vsD-mag-t2.5}
\end{figure}

\begin{figure}[tbh]
    % \centering
    \includegraphics[scale=0.6]{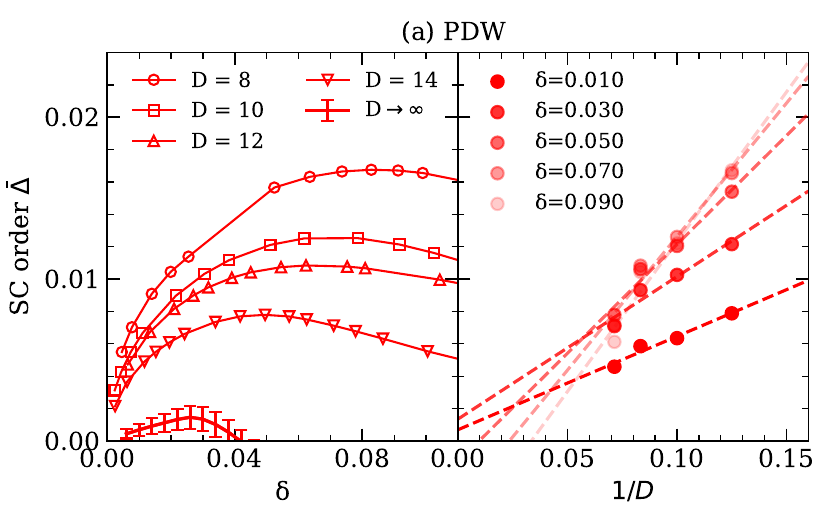}
    \includegraphics[scale=0.6]{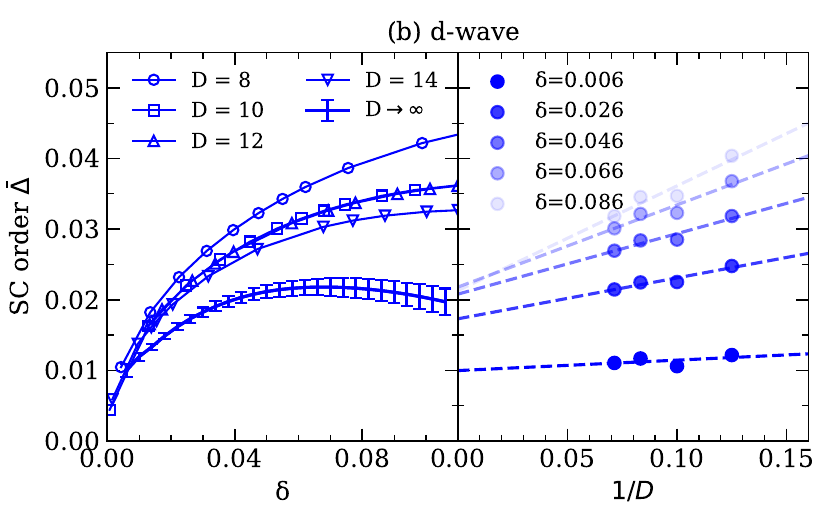}
    \caption{
    $1/D$ scaling of magnitude of NN singlet $|\bar{\Delta}|$ at $t/J = 2.5$ for (a) fluctuating PDW and (b) $d$-wave states measured with $\chi = 64$. }
    \label{fig:vsD-sc-t2.5}
\end{figure}

\section{Fluctuating PDW and \texorpdfstring{$d$}{d}-wave states at \texorpdfstring{$t/J$}{t/J} = 2.5}
\label{app:t2.5}

The general features of the $d$-wave and the fluctuating PDW states when $t/J = 2.5$ are similar to $t/J = 3.0$.
(a) Staggered magnetization vanishes at $\delta \simeq 0.12$ as $D \to \infty$ for the PDW state (Fig. \ref{fig:vsD-mag-t2.5}). 
(b) Although $\Delta_{ij}$ on NN bonds now scales to a nonzero value as $D \to \infty$ when doping $\delta \lesssim 0.04$, it is still very small (Fig. \ref{fig:vsD-sc-t2.5}), especially compared to the $d$-wave value. 
We expect that further optimization on the simple update result will eventually eliminate the SC order in the $D \to \infty$ limit, as in the case of $t/J = 3.0$. 
(c) The PDW state shows the breaking of the $C_4$ symmetry to mere reflections about the diagonal lines, which becomes more evident as doping $\delta$ increases. Meanwhile, the $d$-wave state still has full lattice symmetry. 
(d) The $d$-wave state can be stabilized by turning on a positive NNN hopping $t'$. 

\end{widetext}

\end{document}